\documentclass{stylehepth}

\begin{document}

\title{Supersymmetric intersecting D6-branes and chiral models on the 
$T^6/(Z_4 \times Z_2)$ orbifold}

\author{Gabriele Honecker}

\address{Departamento de F\'{\i}sica Te\'orica C-XI, Facultad de Ciencias, Universidad Aut\'onoma de Madrid, 
Cantoblanco, 28049 Madrid, Spain\\ 
E-mail: gabriele@delta.ft.uam.es}

\maketitle

\abstracts{Intersecting D-brane worlds provide phenomenologically appealing constructions of four dimensional low 
energy string vacua. In this talk, a $Z_4 \times Z_2$ orbifold background is taken into account. It is possible to 
obtain supersymmetric and stable chiral models. In particular, a three generation model with Pati-Salam gauge group 
and no exotic chiral matter is presented.}

\section{Introduction}

In the last four years, intersecting D6-branes have started to play an important role in standard model
constructions from type II string theory. Pioneered by\cite{Blumenhagen:2000wh,Aldazabal:2000cn,Aldazabal:2000dg,Blumenhagen:2000ea,Ibanez:2001nd} (See e.g.\cite{Blumenhagen:2002wn,Uranga:2003pz} and references therein for further works as well as\cite{Angelantonj:2000hi,Larosa:2003mz} for works in the T-dual picture of D9-branes with background fluxes), a large variety 
of non-supersymmetric models with chiral fermions based on intersecting D6-branes wrapping 3-cycles in
an toroidal or orbifold background was constructed and the geometric interpretation of 
gauge and Yukawa couplings as well as proton decay was pointed out, see e.g.\cite{Cremades:2003qj,Cvetic:2003ch,Abel:2003vv,Klebanov:2003my} for more recent work as
 well as\cite{Lust:2003ky} for threshold corrections and\cite{Abel:2003fk} for FCNC . 
In\cite{Cvetic:2001nr} the first supersymmetric 
model with three generations was constructed in an orbifold background $T^6/(Z_2 \times Z_2)$. However, all explicitly known models based on this orbifold suffer from additional `hidden' gauge sectors as well as exotic chiral states charged under the observable gauge group.  On the other hand, the `hidden' gauge sectors can be chosen to be confining and therefore allow for gaugino condensation  which is so far the only known possible source of supersymmetry breaking\cite{Cvetic:2003yd}.     
Subsequently, more supersymmetric chiral models were found in the background $T^6/Z_4$\cite{Blumenhagen:2002gw} 
where fractional D6-branes wrapping exceptional as well as bulk cycles occur. In this case, a three generation model with
only the SM gauge group but some exotic chiral states can be constructed when taking into account the possibility of
 brane recombination.
In this talk, the construction of supersymmetric models on $T^6/(Z_4 \times Z_2)$\cite{Honecker:2003vq} is reformulated in terms of 
homology cycles, and the first three generation model with Pati-Salam gauge group and no exotic chiral states or `hidden' gauge factors is constructed. As in\cite{Blumenhagen:2002gw}, a brane recombination process is required to obtain this result.

\section{3-cycles, RR tadpole cancellation and supersymmetry}

The orientifold $T^6/(\Omega {\cal R} \times Z_4 \times Z_2)$ of type IIA string theory is generated by two orbifold projectors,
\bea 
\Theta: \qquad &(z^1,z^2,z^3) \rightarrow (iz^1,-iz^2,z^3),\nonumber\\
\omega: \qquad &(z^1,z^2,z^3) \rightarrow (z^1,-z^2,-z^3),\nonumber
\eea
where $z^k=x^{2+2k}+ix^{3+2k}$ label the internal complex coordinates. 
The worldsheet parity is  accompanied by 
an antiholomorphic involution,
\begin{equation}
{\cal R}: \qquad z^k \longrightarrow \overline{z}^{\overline{k}}, \qquad k=1,2,3.     \nonumber
\end{equation}
 The orbifold action constrains the geometry to square tori $T^2_1 \times T^2_2$ while 
the involution fixes the orientation of the tori w.r.t. the invariant axes $x^{2+2k}$. The geometric set-up is depicted in figure~1.
\begin{center}
\includegraphics[width=9cm]{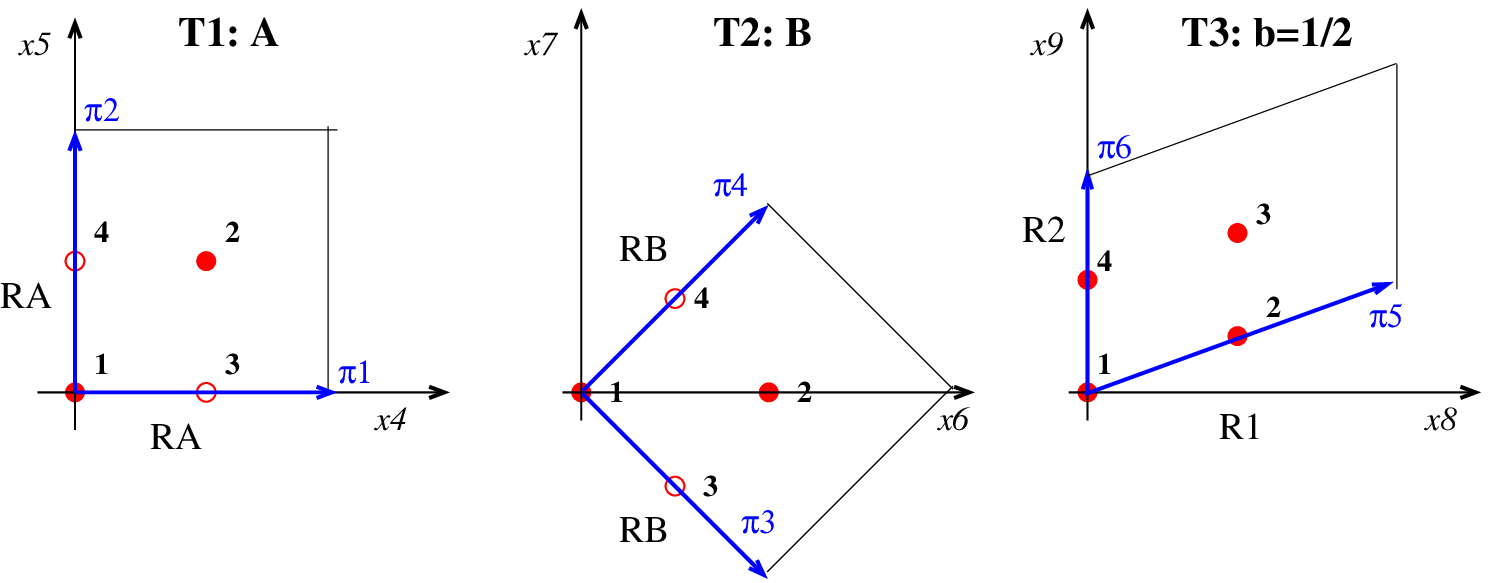}

Figure 1: Cycles and fixed points. Square lattices can have the orientations {\bf A} or {\bf B} w.r.t. the axes $x^{2+2k}$. The third torus $T^2_3$ can be either tilted or untilted.
$Z_4$ fixed points are denoted by filled cycles, additional $Z_2$ fixed points by empty cycles.
\
\vspace{5mm}
\
\end{center}

Four independent 3-cycles of the form  $\rho^{\prime}_1 =(1+\theta +\theta^2+\theta^3)(1+\omega)\pi_{135}$
survive the projection, and the fixed points do not contribute any further 3-cycles as manifestly shown in the Hodge numbers $h_{1,1}=61, h_{2,1}=1$\cite{Klein:2000qw}. The four projected cycles are explicitly given by
\bea 
\rho^{\prime}_1 &=4\left(\pi_{135}-\pi_{245}\right), 
\qquad \rho^{\prime}_3 =4\left(\pi_{136}-\pi_{246}\right),\nonumber \\
\rho^{\prime}_2 &=4\left(\pi_{235}+\pi_{145}\right), 
\qquad  \rho^{\prime}_4 =4\left(\pi_{236}+\pi_{146}\right).\nonumber 
\eea
However, since the only non-vanishing intersection numbers taking into account factors $\frac{1}{4}\cdot\frac{1}{2}$ from the orbifold projector are
$\rho^{\prime}_1 \circ \rho^{\prime}_3=\rho^{\prime}_2 \circ \rho^{\prime}_4=-4$, these cycles do not form an integral basis.
Instead, the basic cycles $\rho_i \equiv \frac{1}{2} \rho^{\prime}_i$ have the required unimodular intersection matrix,
\begin{equation}\label{IntersectionMatrix}
I^{Z_4 \times Z_2} = \left(\begin{array}{cccc}
0 & 0 & -1 & 0\\
0 & 0 & 0 & -1\\
1 & 0 & 0 & 0 \\
0 & 1 & 0 & 0
\end{array}\right).
\end{equation}
Each D6-brane wrapping a factorisable 3-cycle is determined by its wrapping numbers $(n_k,m_k)$ along the cycles 
$\pi_{2k-1},\pi_{2k}$ and has an image under the $Z_4$ rotation,
\begin{equation}
a \Leftrightarrow \left(\begin{array}{cc}
n^a_1, &m^a_1 \\ n^a_2, &m^a_2 \\   n^a_3, &m^a_3
\end{array}\right), \qquad
%%%%%%%%%%%%%%%%%%%%%%%%%%%%%%%%%%
(\Theta a)\Leftrightarrow  \left(\begin{array}{cc}
-m^a_1, &n^a_1 \\ m^a_2, &-n^a_2 \\   n^a_3, &m^a_3
\nonumber 
\end{array}\right).
\end{equation}
The geometry stemming from these two contributions can be compactly rephrased in terms of the basic 3-cycles. A brane $a$ and its image
$(\Theta a)$ wrap the general cycle
\begin{equation}\label{cycleA}
\pi_a = A_a \rho_1 + B_a \rho_2 + C_a \rho_3 + D_a \rho_4
\end{equation}
with the coefficients determined by the wrapping numbers
\bea
A_a &=\left(n^a_1 n^a_2 - m^a_1 m^a_2 \right) n^a_3,
\qquad C_a=\left(n^a_1 n^a_2 - m^a_1 m^a_2 \right)m^a_3, \nonumber\\
B_a &=\left(n^a_1 m^a_2 + m^a_1 n^a_2 \right) n^a_3,
\qquad D_a=\left(n^a_1 m^a_2 + m^a_1 n^a_2 \right) m^a_3. \nonumber
\eea

The discussion up to now is independent of the orientation of the various two tori w.r.t. the axes $x^{2+2k}$. The orientifold symmetry,
however, enforces additional ${\cal R}$ images of the various D-branes which do depend on the choice of the lattice. From now on, we will only consider the lattice orientation {\bf ABb} as displayed in figure~1. The basic cycles then have the following images
with $b=\frac{1}{2},0$ for a tilted or rectangular torus $T^2_3$, respectively,
\bea
\rho_1 &
\stackrel{{\cal R}}{\longrightarrow} \rho_2 - 2b \rho_4,
\qquad 
\rho_3\stackrel{{\cal R}}{\longrightarrow}-\rho_4,\nonumber\\
\rho_2 &
\stackrel{{\cal R}}{\longrightarrow} \rho_1 - 2b \rho_3,
\qquad \rho_4\stackrel{{\cal R}}{\longrightarrow}-\rho_3,\nonumber
\eea
and the image of a general cycle is
\begin{equation}\label{cycleAprime}
\pi_{a'} = B_a \rho_1 + A_a \rho_2 
-\left(D_a+2 b B_a\right) \rho_3 -\left(C_a+2 b A_a\right)   \rho_4.
\end{equation}
Due to the orbifold projection, there exist four different orientifold planes wrapping the following cycles
\bea\label{cyclesO}
\pi_{O6_1} &=\pi_{O6_3} & = \frac{1}{1-b} 
\left[ \left(\rho_1 + \rho_2 \right) -b\left(\rho_3 + \rho_4\right) \right],\\
\pi_{O6_2} &=\pi_{O6_4} & = \rho_3 - \rho_4. \nonumber
\eea
For a more detailed description of the positions of the O6-planes in this case see also\cite{Forste:2000hx}.
Taking also into account the dependence of the number of identical O6-planes, $N_{O6}=2 (1-b)$, on the shape of $T^2_3$, 
the RR tadpole cancellation conditions can now be rephrased in terms of the total homology of the D6-branes compensating for that of the 
O6-planes,  
\begin{equation}
\sum_a N_a \left( \pi_a +\pi_{a'}\right) = 4N_{O6}\pi_{O6},
\end{equation}
with $\pi_{O6}= \sum_{i=1}^4\pi_{O6_i}$ and $N_a$ the number of identical branes $a$. In terms of wrapping numbers, this one equation splits into two non-trivial ones, namely
\bea
\sum_a N_a \left( A_a + B_a \right) &=& 2^4 \nonumber\\
\sum_a N_a \left[\left(C_a - D_a \right) + b \left( A_a - B_a \right)\right]  &=& 2^4(1-b)\nonumber
\eea
matching the RR tadpole cancellation equations given previously\cite{Honecker:2003vq}. At this point, it is important to notice that the 
original statement of traceless matrices $\gamma_{Z_2}$ acting on the Chan-Paton labels of the open string states
and the resulting gauge symmetry breaking 
$U(M) \stackrel{Z_2 \times Z_2}{\longrightarrow} U(M/2)$ is now reformulated in the statement that 
$N$ branes wrapping the cycles $\rho_i \equiv \frac{1}{2} \rho^{\prime}_i$ provide the gauge group $U(N)$.

The complete chiral spectrum as displayed in
table~\ref{tab:gen} can easily be computed from the intersection matrix~(\ref{IntersectionMatrix}) and the 
cycles~(\ref{cycleA},\ref{cycleAprime},\ref{cyclesO}) without the detailed knowledge of the wrapping numbers.
%%%%%%%%%%%%%%%%%%%%%%%%%%%%%%%%%%%%%%%%%%%%%%%%%
\begin{table}[t]
\caption{General chiral spectrum for the  $T^6/(\Omega {\cal R} \times Z_4 \times Z_2)$ orientifold.\label{tab:gen}}
  \begin{center}
\footnotesize
      \begin{tabular}{|c|c|c|} \hline
sector      & multiplicity & representation\\ \hline\hline
$aa$  & & $U(N_a)$\\
 &  $3+[(n^a_1)^2+(m^a_1)^2][(n^a_2)^2+(m^a_2)^2]$ & ${\bf Adj}_a$ \\
$ab$  & $\pi_a \circ \pi_b$ & $({\bf N}_a,\overline{\bf N}_b)$\\
$ab'$ & $\pi_a \circ \pi_{b'}$  & $({\bf N}_a,{\bf N}_b)$\\
%%%%%%%%%%%
$aa'$ &  $\frac{1}{2}\left(\pi_a \circ \pi_{a'}-\pi_a \circ N_{O6}\pi_{O6} \right)$  & ${\bf Sym}_a$\\
%%%%%%%%%%%%%%%%%%%%%%%%%%%%
 &  $\frac{1}{2}\left(\pi_a \circ \pi_{a'}+\pi_a \circ N_{O6} \pi_{O6} \right)$   & ${\bf Anti}_a$\\\hline
      \end{tabular}
  \end{center}
\end{table}
On the contrary, determining the non-chiral part of the spectrum requires a detailed knowledge of the wrapping numbers as can be 
easily seen from the number of adjoint representations arising at intersections of brane $a$ with its image $(\Theta a)$. More details 
concerning the non-chiral spectrum can be found in\cite{Honecker:2003vq}.

Up to this point, the whole construction of intersecting D6-branes with chiral spectrum as displayed in table~\ref{tab:gen} and
satisfying the RR tadpole cancellation conditions does not depend on whether the open string sector is supersymmetric or not.
But since it is well known that non-supersymmetric compactifications require a low string scale in the $TeV$ range whose 
creation through large extra dimensions is ruled out by D6-branes which do not share any common transverse direction, the 
search for phenomenologically interesting spectra is focused on MSSM and supersymmetric GUT model building. 

In the $T^6/(Z_4 \times Z_2)$ case, the only continuous free parameter is the ratio of radii on $T^2_3$. The
condition that a brane $a$ preserves supersymmetry relates this ratio with the wrapping numbers and the discrete quantity 
$b$  parameterising the shape of $T^2_3$, 
\begin{equation}
\frac{R_2}{R_1} \left[\left( C_a + D_a \right) +b \left( A_a + B_a \right)\right] = A_a - B_a.  
\end{equation}
Searching for a general solution to the supersymmetry and RR tadpole cancellation conditions with phenomenologically interesting
chiral spectra turns out to be a highly sophisticated problem. In the following section, a specific three generation model is presented.
Note however, that this simple spectrum is only obtained after a brane recombination process in an ${\cal N}=2$ supersymmetric 
sector has taken place.

\section{A three generation example with Pati-Salam group}

Constraining the search for supersymmetric cycles to those which lie on top of some O6-plane along one two-torus 
leads to six different possibilities, each parameterised by a single non-trivial angle and its negative on the remaining two tori.
From the tadpole cancellation conditions, it is obvious that large wrapping numbers lead to large RR charges which in turn require
the existence of anti-branes in order to cancel the RR tadpoles. Therefore, it is sensible to further constrain the remaining non-trivial angle to be either $0$ or $\frac{\pi}{4}$. Out of the six possibilities with one non-trivial angle, four boil down to branes 
lying on top of some O6-plane and thus only supporting a $Sp(N)$ gauge group. The remaining two non-trivial kinds of cycles giving rise to $U(N)$
gauge factors are given by the following wrapping numbers,
\bea
\left( n^A_1,m^A_1;n^A_2,m^A_2;n^A_3,m^A_3 \right) 
&= \left(1,0;0,1;n^A_3, -\left(\frac{R_1}{R_2}+b\right)n^A_3\right), \nonumber\\
\left( n^B_1,m^B_1;n^B_2,m^B_2;n^B_3,m^B_3 \right) 
&= \left(1,1;1,1;n^A_3, -\left(\frac{R_1}{R_2}+b\right)n^A_3\right),\nonumber
\eea
corresponding to the cycles $\pi_A=\rho_2 - (\frac{R_1}{R_2}+b)\rho_4$ and $\pi_B= 2\pi_A$. As a result,
it is straightforward to compute the following intersection numbers,
\bea
\pi_A \circ \pi_B & =\pi_A \circ \pi_{B'} &=\pi_A \circ \pi_{A'} =\pi_B \circ \pi_{B'}=0,\nonumber\\ 
I^A_{Anti} & =-I^A_{Sym} & =\frac{1}{2}\pi_A \circ \left(N_{O6}\pi_{O6}\right)=2\left((1-b)-\frac{R_1}{R_2}\right),\label{3gen:inter}\\
I^B_{Anti} & =-I^B_{Sym} &=2 I^A_{Anti}.\nonumber
\eea
Due to the correlation of the number of symmetric and antisymmetric representation in~(\ref{3gen:inter}), starting with this approach it is obviously impossible to obtain some $SU(5)$ GUT model without chiral states transforming in the symmetric representation. $SO(10)$ GUTs are not accessible either in this set-up since there is no 
way to reproduce the spinor representation starting from a perturbative D-brane configuration.

Fixing the shape of $T^2_3$ such that no symmetric representation occurs, i.e. $\frac{R_1}{R_2}=b$, it is possible to construct a 
model with Pati-Salam gauge group and 2+1 generations with the choice of stacks and wrapping numbers displayed in table~\ref{tab:3gen}.
%%%%%%%%%%%%%%%%%%%%%%%%%%%%%%%%%%%%%%%%%%%%%%%%%
\begin{table}[t]
\caption{Wrapping numbers for a three generation model with Pati-Salam group.\label{tab:3gen}}
  \begin{center}
\footnotesize
      \begin{tabular}{|c|c||c|c|c||c|} \hline
brane & $N$ & $(n_1,m_1)$ & $(n_2,m_2)$ & $(n_3,m_3)$ & cycle\\ \hline\hline
$A$ & 4 & (1,0) & (0,1) & (1,-1) & $(\rho_2 - \rho_4)$\\
$B$ & 4 & (1,1) & (1,1) & (1,-1) & $2(\rho_2 - \rho_4)$\\
$C$ & 2 & (1,0) & (1,1) & (2,-1) & $2(\rho_1+\rho_2)-(\rho_3+\rho_4)$\\
$D$ & 2 & (1,1) & (1,0) & (0,-1) & $(\rho_3 - \rho_4)$\\\hline
      \end{tabular}
  \end{center}
\end{table}
Since this brane set-up fulfills the RR tadpole cancellation conditions, no hidden gauge sector will appear. The initial gauge group
is $U(4)_A \times U(4)_B \times Sp(2)_C \times Sp(2)_D$. Due to $SU(2) \simeq Sp(2)$ the two latter factors can be identified with  
$SU(2)_L   \times SU(2)_R$. The intersections of the brane orbits $A$ and $B$ provide two hypermultiplets in the bifundamental representation $({\bf 4}_A,\overline{\bf 4}_B)+h.c.$ whose vev's can be chosen such as to fulfill the F- and D-flatness conditions while recombining the two kinds of branes into a single non-factorisable one. The gauge group can further be broken down to $SU(3)_C \times SU(2)_L \times
U(1)_Y \times U(1)_X$ by performing two parallel displacements of branes: separating the $U(4)$ stack on $T^2_3$ into
$[SU(3) \times U(1)_1] \times U(1)_2$ gives the massless gauge factor $SU(3) \times U(1)_{B-L}$ with 
$ U(1)_{B-L}=\frac{1}{3}(U(1)_1 -3 U(1)_2)$ while the orthogonal linear combination becomes massive by the Green Schwarz mechanism.
Finally, breaking $ SU(2)_R \times U(1)_{B-L}$ by parallely displacing the stack $D$ off the O6-plane e.g. on $T^2_3$ provides the 
standard model hypercharge as well as an additional massless Abelian gauge factor, $U(1)_{Y/X}  = U(1)_{B-L} \pm U(1)_R$.
The choice of the sign distinguishes the two different possibilities of identifying right-handed up- and down-type quarks and leptons.
The complete breaking mechanism is displayed in diagram~\ref{BreakingPattern}.
\bea
& \underbrace{U(4)_A \times U(4)_B}   \times SU(2)_R   \times SU(2)_L \nonumber\\
& \qquad \downarrow   \nonumber \mbox{Brane recombination} \label{BreakingPattern}\\
& U(4)  \times SU(2)_R   \times SU(2)_L \nonumber\\
& \qquad \downarrow   \mbox{Parallel displacement \& GS mechanism} \\
& \overbrace{SU(3) \times \underbrace{U(1)_{B-L}} \times U(1)_{massive}} \times \underbrace{SU(2)_R} \times SU(2)_L \nonumber \\
& \qquad \downarrow   \nonumber \mbox{Parallel displacement}\\
& SU(3)  \times SU(2)_L  \times U(1)_Y  \times U(1)_X    \times U(1)_{massive}\nonumber
\eea
The chiral spectrum for each step of the symmetry breaking pattern is displayed in diagram~\ref{ChiralSpectrum}.
\bea
& \Bigl[({\bf 4}_A, {\bf 2}_L) +({\bf \overline{4}}_A, {\bf 2}_R) \Bigr]
+ 2 \Bigl[({\bf 4}_B, {\bf 2}_L) +({\bf \overline{4}}_B, {\bf 2}_R) \Bigr] \nonumber \\
& \qquad \downarrow   \nonumber \\
& 3 \Bigl[({\bf 4}, {\bf 2}_L)\Bigr. + \Bigl.({\bf \overline{4}}, {\bf 2}_R) \Bigr] \nonumber \\
& \qquad \downarrow    \label{ChiralSpectrum} \\
& 3 \Bigl[({\bf 3}, {\bf 2}_L)_{1/3} +({\bf 1}, {\bf 2}_L)_{-1} 
+ ({\bf \overline{3}}, {\bf 2}_R)_{-1/3}  +({\bf 1}, {\bf 2}_R)_{1}\Bigr] \nonumber \\
& \qquad \downarrow   \nonumber \\
& 3 \Bigl[({\bf 3}, {\bf 2}_L)_{(1/3,1/3)} +({\bf 1}, {\bf 2}_L)_{(-1,-1)}
+({\bf \overline{3}})_{(-4/3,2/3)}  + ({\bf \overline{3}})_{(2/3,-4/3)} 
+({\bf 1})_{(2,0)}+({\bf 1})_{(0,2)}\Bigr] \nonumber 
\eea
In this specific model, the sector providing the standard model Higgs particles consists of 
two half-hypermultiplets in $({\bf 2}_L, {\bf 2}_R)$ of $SU(2)_L \times SU(2)_R$ at the initial stage and decomposes into
two half-hypermultiplets transforming as $({\bf 2}_L)_{(1,-1)}+({\bf 2}_L)_{(-1,1)}$ 
under $SU(2)_R \times U(1)_{B-L}  \rightarrow U(1)_Y  \times U(1)_X$. 
Since the electro weak symmetry breaking is expected to occur only after supersymmetry breaking, we terminate the discussion of the 
symmetry breaking pattern at this point. It should, however, be emphasized, that due to the symplectic gauge factors, half-hypermultiplets occur in the non-chiral sectors, whereas in a more general set-up, unitary gauge factors supply for more Higgs candidates sitting in full ${\cal N}=2$ hypermultiplets.

Finally, let us conclude by pointing out that the set-up in table~\ref{tab:3gen} fulfills the conditions 
on the wrapped cycles for stringy gauge coupling unification\cite{Blumenhagen:2003jy}. But due to the brane recombination process,
a non-factorisable stack of branes occurs which hinders from calculating couplings and threshold corrections since the known results 
using conformal field theory apply only to factorisations into two-tori.

\section*{Acknowledgments}
%I am grateful to Daniel Cremades, Luis Ib\'a\~{n}ez, Fernando Marchesano as well as Lars G\"orlich, Tassilo Ott and especially Ralph Blumenhagen and Angel Uranga.
This work is supported by the EC under RTN program number HPRN-CT-2000-00148.

\end{document}